\documentclass{PoS}

\usepackage[utf8]{inputenc}
\usepackage{mathtools}
\usepackage{cleveref}

\usepackage[numbers, square, comma, sort&compress]{natbib}

\usepackage{listings}
\lstdefinestyle{Common}
{
    basicstyle=\small\ttfamily\null,
    numbers=none,
    numbersep=1em,
    frame=none,
    framesep=\fboxsep,
    framerule=\fboxrule,
    xleftmargin=\dimexpr\fboxsep+\fboxrule,
    xrightmargin=\dimexpr\fboxsep+\fboxrule,
    breaklines=true,
    breakindent=0pt,
    tabsize=5,
    columns=flexible,
    showstringspaces=false,
    captionpos=b,
    abovecaptionskip=0\smallskipamount,
}
\lstdefinestyle{Mathematica}
{
    style=Common,
    language={[LaTeX]TeX},
}
\lstdefinestyle{CodeSample}
{
    style=Mathematica
}
\newcommand{\Math}[1]
{\lstinline[style=Mathematica,breaklines=false,basicstyle=\small \ttfamily\null]~#1~}
\lstnewenvironment{CodeSample}[1][]
{\vspace{-3mm}\lstset{style=CodeSample,#1}}
{}

\def\d{{\rm d}}

\usepackage{tikz}
\usetikzlibrary{positioning}
\usetikzlibrary{arrows}
\usetikzlibrary{decorations.pathmorphing}
\usetikzlibrary{decorations.pathreplacing}
\usetikzlibrary{decorations.markings}
\usepackage{etoolbox}
\usetikzlibrary{backgrounds}
\usetikzlibrary{calc}

\pgfarrowsdeclare{varstealth}{varstealth}
 {
  \pgfarrowsleftextend{-5.59\pgflinewidth}
  \pgfarrowsrightextend{4.5\pgflinewidth}}
 {
  \pgfpathmoveto{\pgfpoint{4.5\pgflinewidth}{0pt}} 
  \pgfpathlineto{\pgfpoint{-0.63\pgflinewidth}{1.09\pgflinewidth}} 
  \pgfpathlineto{\pgfpoint{-5.48\pgflinewidth}{3.06\pgflinewidth}} 
  \pgfpathlineto{\pgfpoint{-5.59\pgflinewidth}{3.0\pgflinewidth}}
  \pgfpathlineto{\pgfpoint{-3.95\pgflinewidth}{0pt}} 
  \pgfpathlineto{\pgfpoint{-5.59\pgflinewidth}{-3.0\pgflinewidth}}
  \pgfpathlineto{\pgfpoint{-5.48\pgflinewidth}{-3.06\pgflinewidth}}
  \pgfpathlineto{\pgfpoint{-0.63\pgflinewidth}{-1.09\pgflinewidth}}
  \pgfusepathqfill
 }
 
\tikzset{
	cut/.style={line width=0.3mm,draw=black, 
		postaction={decorate},
    		decoration={markings,mark=at position .5 with {\arrow[red,line width=0.35mm]{|}}}},
	massless/.style={line width=0.3mm},
	directed/.style={line width=0.3mm,postaction={decorate},
    		decoration={markings,mark=at position .6 with {\arrow[line width=0.3mm]{varstealth}}}},
	external/.style={line width=0.3mm,postaction={decorate},
    		decoration={markings,mark=at position .75 with {\arrow[line width=0.3mm]{varstealth}}}},
}

\newcommand\Spaa[2]{\langle#1\,#2\rangle}						
\newcommand\Spab[3]{\langle#1|\hspace{0.2mm}#2\hspace{0.2mm}|#3]}						

\title{{\sc MultivariateResidues} -- a Mathematica package for computing multivariate residues}
\ShortTitle{{\sc MultivariateResidues:} a Mathematica package for computing multivariate residues}

\author{\speaker{Robbert Rietkerk}\\
       Institute for Theoretical Particle Physics, KIT, \\
       Wolfgang-Gaede-Strasse 1, 76128 Karlsruhe, Germany\\
       E-mail: \email{Robbert.Rietkerk@kit.edu}}

\author{Kasper J. Larsen\\
        School of Physics and Astronomy, University of Southampton,\\ 
        Highfield, Southampton, SO17 1BJ, United Kingdom\\
        E-mail: \email{Kasper.Larsen@soton.ac.uk}}

\abstract{
We present the Mathematica package {\sc MultivariateResidues}, which allows for the efficient evaluation of multivariate residues based on methods from computational algebraic geometry.
Multivariate residues appear in several contexts of scattering amplitude computations.
Examples include applications to the extraction of master integral coefficients from maximal unitarity cuts, the construction of canonical bases of loop integrals and the construction of tree amplitudes from scattering equations.
}

\FullConference{13th International Symposium on Radiative Corrections (Applications of Quantum Field Theory to Phenomenology)\\
                 25-29 September, 2017 \\
                 St. Gilgen, Austria}

\begin{document}

\section{Introduction}
Scattering amplitudes are essential ingredients to theoretical studies of scattering processes at particle colliders.
It is important to calculate scattering amplitudes with sufficient accuracy, which is achieved by taking higher-order radiative corrections into account. 
The resulting higher-order amplitudes depend on increasingly many internal and/or external variables and thus naturally require methods of multivariate calculus.
In particular, multivariate residues make their appearance in the context of two-loop amplitude calculations.

Multi-loop amplitudes are systematically calculated in terms of Feynman diagrams. 
The large set of corresponding Feynman integrals are usually reduced to a much smaller set of master integrals by solving a system of integration-by-parts (IBP) identities among the Feynman integrals, a process known as IBP reduction \cite{Chetyrkin:1981qh,Laporta:2000dc}.
The coefficients of master integrals produced by the IBP reduction are rational functions of the external invariants of the scattering amplitude and the space-time dimension $d$.
As the number of external invariants increases, expressions for integral coefficients tend to grow dramatically in size, in particular at intermediate stages of the reduction.
As a result, IBP reduction can become a serious bottleneck in the computation of scattering amplitudes.
A direct way of obtaining integral coefficients, bypassing the IBP reduction, would therefore be of practical interest.

Generalised unitarity has been used with much success in calculations of one-loop amplitudes with many external invariants \cite{Bern:1994zx,Ellis:2009zw}.
The feat of bypassing IBP reductions is achieved in its generalisation to two loops \cite{Kosower:2011ty,CaronHuot:2012ab}.
In generalised unitarity, the coefficients of a loop amplitude decomposed in a basis of integrals are obtained as residues of products of tree amplitudes. 
As a result, this method leads one to consider multivariate residues.

Multivariate residues appear in other contexts as well.
For instance, they can be used to calculate leading singularities of Feynman integrals, in search of canonical bases of integrals \cite{Henn:2013pwa}.
Multivariate residues also play a central role in the Grassmannian formulation of the S-matrix \cite{ArkaniHamed:2009dn}.
Furthermore, multivariate residues provide an efficient means to localise amplitudes computed in the Cachazo-He-Yuan formalism to the solutions of scattering equations \cite{Cachazo:2013hca,Sogaard:2015dba}.

Multivariate residues can be non-trivial to evaluate in practice.
Nevertheless, implementations of their evaluation have not been publicly available. 
In this talk we discuss the recently developed Mathematica package {\sc MultivariateResidues} \cite{Larsen:2017aqb} for evaluating multivariate residues.
We conclude by providing three example applications.

\section{Definition of multivariate residues}
\label{sec:definition_residue}

The setup of our work is as follows. We consider a given $n$-form
\begin{align}
\Omega = \frac{h(z) \hspace{0.4mm} \d z_1\wedge\cdots\wedge \d z_n}{f_1(z)\cdots f_n(z)}~,
\label{eq:n_form}
\end{align}
where $h(z)$ and $f_{i}(z)$ are polynomials of the complex variables $z = (z_1, \dotsc, z_n)$.
A pole of $\Omega$ is defined as a point $p \in \mathbb{C}^n$ where $f(z) = (f_1(z), \dotsc, f_n(z))$ has an isolated zero.
The multivariate residue of $\Omega$ is then defined as \cite{GriffithsHarris},
\begin{align}
\mathop{\mathrm{Res}}_{\{f_1,\dots,f_n\}, \hspace{0.6mm} p}(\Omega)
=\frac{1}{(2\pi i)^n}\oint_{\Gamma}
\Omega~,
\label{eq:residue_definition}
\end{align}
where the contour $\Gamma$ encircles each denominator factor,
\begin{align}
\Gamma = \{z \in \mathbb{C}^n : |f_i(z)| = \epsilon_i ~~~\forall~~ i=1,\dotsc,n \}~,
\label{eq:contour}
\end{align}
with the $\epsilon_i$ infinitesimal.
The contour is oriented by the condition $\d (\mathrm{arg} \hspace{0.6mm} f_1) \wedge \cdots \wedge \d (\mathrm{arg} \hspace{0.6mm} f_n) \geq 0$.

The $n$-form in \cref{eq:n_form} has precisely as many denominator factors as complex variables. 
It suffices to discuss multivariate residues of such forms, because other situations can be reduced to this case.
Should there be \emph{less} denominator factors than integration variables, then the present definition of multivariate residues is not immediately applicable. 
However, one can still take iterated residues with respect to subsets of the $n$ complex variables, and for each residue apply the definition of the multivariate residue.
Should there be \emph{more} denominator factors than integration variables, then one can always partition the set of denominator factors into precisely $n$ factors.
In general, each partitioning leads to a different value for the residue, 
since the residue in eqs.~(\ref{eq:residue_definition})--(\ref{eq:contour}) is defined in terms of the denominator factors $f_i(z)$.
The fact that the multivariate residue is not uniquely specified by the location of the pole is an inherent feature of multivariate residues.
The geometric picture is that in the multivariate case it becomes possible to encircle a given pole with contours which are inequivalent in the sense that one contour cannot be continuously deformed into the other contour without crossing a singular surface of the integrand, as explained in the appendix of ref.~\cite{Larsen:2017aqb}.

\section{Evaluation of multivariate residues}
\label{sec:evaluation_residue}

We will now discuss how to evaluate the multivariate residue in \cref{eq:residue_definition} in practice.
Let's start with the simplest case, for which it is sufficient to change variables $z \to w = f(z)$.
Indeed, if the associated Jacobian,
\begin{align}
\mathrm{Jac}(p)\equiv \det_{i,j}\left(  \frac{\partial f_i}{\partial z_j}  \right)\bigg|_{z = p}~,
\label{eq:jacobian}
\end{align}
is non-vanishing, then the residue is said to be non-degenerate and evaluates to
\begin{align}
\mathop{\mathrm{Res}}_{\{ f_1,\dots,f_n \},p}(\Omega)
= \frac{h(p)}{\mathrm{Jac}(p)}~.
\label{eq:simple_residue}
\end{align}
On the other hand, if the Jacobian in \cref{eq:jacobian} vanishes, then the residue is said to be degenerate, and a different computational strategy is needed. 

Conceptually the simplest way to evaluate degenerate multivariate residues is by performing a transformation of denominator factors (rather than a transformation of variables).
A theorem in algebraic geometry, see chapter 5 in ref.~\cite{GriffithsHarris}, states that for certain linear combinations of the denominator factors, namely $g_i (z) = \sum_{j=1}^n A_{ij} (z) f_j(z)$ with locally holomorphic $A_{ij}(z)$, the following transformation formula holds,
\begin{align}
\mathop{\mathrm{Res}}_{\{ f_1, \ldots, f_n \}, \hspace{0.3mm} p} \hspace{-0.2mm}\left(
\frac{h(z) \hspace{0.4mm} \d z_1 \wedge \cdots \wedge \d z_n}{f_1 (z) \cdots f_n (z)} \right)
\hspace{1mm}=\hspace{1mm} \mathop{\mathrm{Res}}_{\{ g_1, \ldots, g_n \}, \hspace{0.3mm} p}
 \hspace{-0.2mm}\left(
\frac{h(z) \hspace{0.1mm} \d z_1 \wedge \cdots \wedge \d z_n}
{g_1 (z) \cdots g_n (z)} \, \det A(z) \right)~.
\label{eq:transformation_formula}
\end{align}
The key idea is to change to univariate denominator factors $g_i(z) = g_i(z_i)$, which can be constructed systematically (although computationally expensive) via Gr\"obner basis computations, after which the multivariate residue factorises into a product of univariate residues.

A more sophisticated method of evaluating degenerate multivariate residues exploits the fact that the residue defines a non-degenerate inner product on the quotient ring consisting of all polynomials in the variables $z_1,\dotsc,z_n$ with coefficients in $\mathbb{C}$ modulo the zero-dimensional ideal generated by the denominator factors $f(z)$.
After decomposing the numerator $h(x)$ in the canonical basis of the quotient ring, and the constant $1$ in the dual basis with respect to the residue map, the \emph{global residue} is computed as the inner product of the coefficient vectors associated with the two decompositions.
Local residues are extracted by multiplying the numerator $h(x)$ with partition-of-unity polynomials. 
For a more detailed description of this method we refer to ref.~\cite{Larsen:2017aqb}.

An implementation of the above-mentioned methods for calculating multivariate residues has been made publicly available in the Mathematica package {\sc MultivariateResidues}.
In line with the discussion in \cref{sec:definition_residue}, the new function \Math{MultivariateResidue} requests separate specifications of the numerator function $h(z)$ and the denominator factors $f(z) = (f_1(z),\dotsc,f_n(z))$.
If necessary, a partitioning of $m>n$ denominator factors into exactly $n$ subsets should be performed by the user in advance.
It should also be emphasised that the functions $h(z)$ and $f(z)$ are restricted to be polynomials in the complex variables. 
Gamma functions of $z$, for instance, are beyond our scope.
An extensive manual for {\sc MultivariateResidues} can be found in ref.~\cite{Larsen:2017aqb}.
A simple example of a multivariate residue computation is the following:
\begin{CodeSample}
<< "MultivariateResidues`"
h = z1;   f1 = z2;   f2 = (a1*z1 + a2 z2)*(b1*z1 + b2*z2);
MultivariateResidue[h, {f1, f2}, {z1->0, z2->0}]
Out: -1/(a1*b1)
\end{CodeSample}

\section{Applications}
\label{sec:examples}
In the following three subsections we give three examples of the application of multivariate residues in the context of scattering amplitude calculations.

\subsection{Master integral coefficients}
\label{sec:maximal_unitarity}

Our first example is about the extraction of master integral coefficients.
Let us start by discussing the context of unitarity in amplitude calculations.
It is well-known that a one-loop amplitude can be decomposed in terms of a small set of one-loop master integrals \cite{Bern:1994zx}. 
Schematically, 
\begin{align}
\mathcal{A}^{\text{one-loop}} = \sum_{k} \, c_k \, I_k + \text{rational terms}~.
\label{eq:amplitude_decomposition}
\end{align}
Moreover, the integral coefficients $c_k$ are cut-constructible in four dimensions.
A famous example is the coefficient of the one-loop box integral \cite{Britto:2004nc}, which can be extracted from \cref{eq:amplitude_decomposition} by replacing the loop-momentum integration along the real slice with the contour $T^4_\epsilon = \{  \ell \in \mathbb{C}^4 : |p_i^2(\ell)|=\epsilon_i\}$.
In fact, all integral coefficients in \cref{eq:amplitude_decomposition} can be written in terms of non-degenerate multivariate residues, which are evaluated straightforwardly by a change of variables, cf. \cref{eq:simple_residue}.

For two-loop amplitudes, the unitarity approach can be pursued by considering maximal cuts of two-loop integrals.
It has been shown that the coefficients of double-box integrals, to the leading order in $(d-4)$, are indeed cut-constructible \cite{Kosower:2011ty}.
While the latter computation involved non-degenerate multivariate residues,  proceeding towards sub-topologies of the double box leads naturally to degenerate multivariate residues, whose evaluation require the methods from algebraic geometry discussed in \cref{sec:evaluation_residue}.

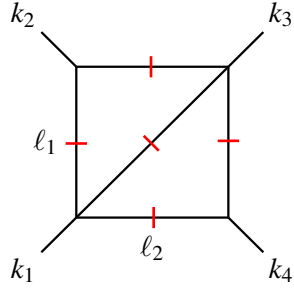
\begin{figure}
\centering
\begin{tikzpicture}[scale=2]
\draw[cut] (0,0) -- (0,1) node[midway,left=1mm] {$\ell_1$}; 
\draw[cut] (0,1) -- (1,1); 
\draw[cut] (1,1) -- (1,0); 
\draw[cut] (1,0) -- (0,0) node[midway,below=1mm] {$\ell_2$} ; 
\draw[cut] (0,0) -- (1,1); 
\draw[massless] (0,0) -- +(-0.23,-0.23) node[below left=-1mm] {$k_1$};  
\draw[massless] (0,1) -- +(-0.23,0.23) node[above left=-1mm] {$k_2$};
\draw[massless] (1,1) -- +(0.23,0.23) node[above right=-1mm] {$k_3$};  
\draw[massless] (1,0) -- +(0.23,-0.23) node[below right=-1mm] {$k_4$};  
\end{tikzpicture}
\caption{The maximally cut slashed-box integral.}
\label{fig:slashed-box}
\end{figure}

Consider, for instance, the slashed-box integral. 
Its maximal cut, illustrated in \cref{fig:slashed-box}, depends on $2 \times 4 - 5 = 3$ complex variables ($z_1,z_2,z_3$).
Localising the remaining three variables yields degenerate multivariate residues.
A typical example is given by
\begin{align}
\mathop{\mathrm{Res}}_{\{f_1,f_2,f_3\}, \hspace{0.6mm} (0,1,0)}\left(\frac{\d z_1 \wedge \d z_2 \wedge \d z_3}{f_1 \, f_2 \, f_3}\right)
= \frac{1+\chi}{\chi}~,
\end{align}
where $f_1 = z_1 (1-z_1-z_2),\, f_2 = z_2 \, z_3,\, f_3 = (1 - z_1 - z_2 - z_1 \chi + z_1 z_3 \chi)$, and $\chi = t/s$ is the usual ratio of Mandelstam invariants.
This example is one of many residues, obtained by different partitions of denominator factors into three functions $f_1, f_2, f_3$.
As a result, the computation of the slashed box provides many test-cases for the {\sc MultivariateResidues} package.
The resulting set of $6395$ residues in three complex variables was evaluated (on a single core on a standard laptop) in about ten minutes, averaging to approximately one-tenth of a second per residue.

\subsection{Canonical basis of integrals}
\label{sec:canonical_basis}

Our second example concerns master integrals.
The computation of master integrals benefits from the freedom to choose any particular basis of integrals, because the form of the differential equations for the basis integrals depends greatly on that choice of basis.
In particular, differential equations in canonical form ($\epsilon$-factorised and Fuchsian) are trivial to solve.
Finding an associated canonical basis of integrals is therefore an important problem.
This is witnessed by the recent appearance of several public programs, Fuchsia \cite{Gituliar:2017vzm}, epsilon \cite{Prausa:2017ltv} and Canonica \cite{Meyer:2017joq}, which address aspects of this problem. 
A different strategy for finding a canonical basis follows an older idea of constructing Feynman integrals with unit leading singularity \cite{ArkaniHamed:2012nw}.
A basis of such integrals is conjectured to be a canonical basis \cite{Henn:2013pwa}.
For instance, the leading singularity of the one-loop box integral is $1/(st)$, and one finds that indeed $s\,t\,I_{\text{box}}$ is a suitable canonical master integral.
Since leading singularities are essentially multivariate residues of Feynman integrands, they can be calculated with {\sc MultivariateResidues}.

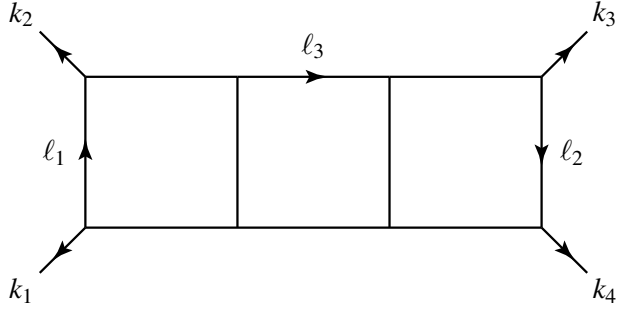
\begin{figure}
\centering
\begin{tikzpicture}[scale=2]
\draw[directed] (0,0) -- (0,1) node[midway,left=1mm] {$\ell_1$}; 
\draw[massless] (1,0) -- (1,1); 
\draw[massless] (2,0) -- (2,1); 
\draw[directed] (3,1) -- (3,0) node[midway,right=1mm] {$\ell_2$}; 
\draw[massless] (0,1) -- (1,1);
\draw[directed] (1,1) -- (2,1) node[midway,above=1mm] {$\ell_3$};  
\draw[massless] (2,1) -- (3,1); 
\draw[massless] (0,0) -- (3,0); 
\draw[external] (0,0) -- +(-0.3,-0.3) node[below left=-1mm] {$k_1$};  
\draw[external] (0,1) -- +(-0.3,0.3) node[above left=-1mm] {$k_2$};
\draw[external] (3,1) -- +(0.3,0.3) node[above right=-1mm] {$k_3$};  
\draw[external] (3,0) -- +(0.3,-0.3) node[below right=-1mm] {$k_4$};  
\end{tikzpicture}
\caption{The massless planar triple-box integral.}
\label{fig:triple-box}
\end{figure}

As an illustration of the computation of leading singularities, consider the planar triple-box integral in \cref{fig:triple-box} with all particles massless, which was first calculated in ref.~\cite{Henn:2013fah}.
We adopt the momentum flow conventions shown in \cref{fig:triple-box} and denote an arbitrary triple-box integral with numerator $N$ by $I_{\text{triple-box}}(N)$.
In this notation, the three triple-box integrals used as canonical master integrals in ref.~\cite{Henn:2013fah} are,
\begin{align}
I_{\text{triple-box}}(1)\,,~~~ 
I_{\text{triple-box}}\big((\ell_1-k_2-k_3)^2\big)\,,~~~ 
I_{\text{triple-box}}\big((\ell_3-k_3)^2\big)~.
\label{eq:master_integrals}
\end{align}
These integrals were selected for their unit leading singularity properties, which can be confirmed in the following way.
First, it is convenient to parametrise the loop momenta as \cite{Sogaard:2013fpa},
\begin{align}
\ell_1^\mu = 
L_{123}^\mu(\alpha_1,\dotsc,\alpha_4) &= 
\alpha_1 k_1^\mu +
\alpha_2 k_2^\mu +
\frac{\alpha_3}{2} \frac{\Spaa{2}{3}}{\Spaa{1}{3}} \Spab{1}{\sigma^\mu}{2} +
\frac{\alpha_4}{2} \frac{\Spaa{1}{3}}{\Spaa{2}{3}} \Spab{2}{\sigma^\mu}{1}~,
\end{align}
and similarly $\ell_2^\mu = L_{234}(\beta_1,\dotsc,\beta_4)$, $\ell_3^\mu = L_{341}(\gamma_1,\dotsc,\gamma_4)$.
Next, one solves the constraint that all propagators are on shell in terms of the $\alpha_i,\beta_i,\gamma_i$. 
There are fourteen distinct solutions \cite{Badger:2012dv}.
For each solution, residues associated with the on-shell propagators are evaluated, leaving rational functions of two variables. 
Those rational functions can in turn be localised to their poles, and it turns out that these residues are degenerate \cite{Sogaard:2013fpa}. 
With {\sc MultivariateResidues} it is straightforward to calculate such degenerate multivariate residues.
As a result, one finds that the integrals in \cref{eq:master_integrals} have constant leading singularities and are therefore suitable elements of a canonical basis of integrals.
Thus, the package {\sc MultivariateResidues} can provide a small step in finding canonical bases of integrals.

\subsection{Tree-level amplitudes from scattering equations}
Our third example is the application of multivariate residues in the context of the Cachazo-He-Yuan (CHY) scattering equations, which describe scattering in arbitrary spacetime dimension \cite{Cachazo:2013hca}.
The scattering equations relate the kinematics of $n$ massless particles to $n$ points $z_1,\dotsc,z_n$ on a Riemann sphere,
\begin{align}
\sum_{j=1, \, j \neq i}^{n} \,\frac{s_{ij}}{z_i - z_j} = 0 ~, ~~~ i \in \{1,\dotsc,n\}~.
\end{align}
The scattering equations are invariant under M\"obius transformations, which leaves $(n-3)!$ inequivalent solutions.
The CHY formula for a tree-level amplitude for $n$-particle scattering is given by a contour integral that localises a suitable integrand to all inequivalent solutions of the scattering equations.
Although the CHY formula is elegant and compact, the instruction to sum over a factorially growing number of residues makes it difficult to compute in practice, already at relatively low multiplicities. 
Moreover, for $n>5$ the solutions become irrational, while the final sum of all residues is a simpler rational function \cite{Weinzierl:2014vwa}.
In ref.~\cite{Sogaard:2015dba} it was noted that these two difficulties are circumvented by direct evaluation of the global residue mentioned toward the end of \cref{sec:evaluation_residue}.

As a simple illustration, consider a five-scalar amplitude in $\phi^3$-theory \cite{Sogaard:2015dba}.
Due to M\"obius invariance we can fix $(z_1,z_2,z_5)=(\infty,1,0)$.
The remaining two variables $z_3,z_4$ must be localised to the solutions of the scattering equations, which is done efficiently by taking the global residue.
In {\sc MultivariateResidues}, precisely such a computation is performed by
\vspace{-2mm}
\begin{CodeSample}
MultivariateResidue[h, {f1,f2}, {z1,z2}, {GlobalResidue}, 
                                                  Method -> "QuotientRingDuality"]
\end{CodeSample}
This approach reproduces the well-known tree-level five-scalar amplitude.
We emphasise that the instruction \Math{\{GlobalResidue\}} prompts the global residue to be calculated directly, without summing over individual residues in the intermediate stage.
Following the method of ref.~\cite{Sogaard:2015dba}, multivariate residues thus allow for the efficient calculation of tree-level scattering amplitudes.

\setlength{\bibsep}{0.0pt}
\bibliographystyle{JHEP}
\bibliography{PoS_MultivariateResidues}

\end{document}